\documentclass{article}

\usepackage{graphicx}

\usepackage{amsmath,latexsym}

\begin{document}

\setcounter{page}{1}

\pagestyle{plain}

\begin{center}
\Large{\bf Observational Status of Tachyon Natural Inflation and Reheating}\\
\small \vspace{1cm} { Narges
	Rashidi}$^{a,}$\footnote{n.rashidi@umz.ac.ir}\,, \quad { Kourosh
	Nozari}$^{a,b,}$\footnote{knozari@umz.ac.ir}
\quad and \quad {{\O}yvind Gr{\o}n}$^{c,}$\footnote{oyvind.gron@hioa.no} \\
\vspace{0.5cm} $^{a}$Department of Physics, Faculty of Basic
Sciences,
University of Mazandaran,\\
P. O. Box 47416-95447, Babolsar, IRAN\\
$^{b}$ Research Institute for Astronomy and Astrophysics of Maragha (RIAAM),\\
P. O. Box 55134-441, Maragha, Iran\\
$^{c}$ Oslo and Akershus University College of Applied Sciences,
Faculty of Technology, Art and Design,\\ PbB4 St. Olavs Plass,
NO-0130 Oslo, Norway
\end{center}

\begin{abstract}
We study observational viability of Natural Inflation with a tachyon
field as inflaton. By obtaining the main perturbation parameters in
this model, we perform a numerical analysis on the parameter space
of the model and in confrontation with $68\%$ and $95\%$ CL regions
of Planck2015 data. By adopting a warped background geometry, we
find some new constraints on the width of the potential in terms of
its height and the warp factor. We show that the Tachyon Natural
Inflation in the large width limit recovers the tachyon model with a
$\phi^{2}$ potential which is consistent with Planck2015
observational data. Then we focus on the reheating era after
inflation by treating the number of e-folds, temperature and the
effective equation of state parameter in this era. Since it is
likely that the value of the effective equation of state parameter
during the reheating era to be in the range $0\leq \omega_{eff}\leq
\frac{1}{3}$, we obtain some new constraints on the tensor to scalar
ratio, $r$, as well as the e-folds number and reheating temperature
in this Tachyon Natural Inflation model. In particular, we show that
a prediction of this model is $r\leq\frac{8}{3}\,\delta_{ns}$, where
$\delta_{ns}$ is the scalar spectral tilt, $\delta_{ns}=1-n_{s}$. In
this regard, given that from the Planck2015
data we have $\delta_{ns}=0.032$ (corresponding to $n_{s}=0.968$), we get $r\leq 0.085$.
\end{abstract}
\newpage

\section{Introduction}

After introduction of ``cosmological inflation" by Guth in
1981~\cite{Gut81}, the idea of the rolling scalar field driving the
dynamics of the early inflationary expansion was introduced by
Linde~\cite{Lin82} and Albrecht and Steinhardt~\cite{Alb82}.
Inflation models predict some small inhomogeneities (caused by the
quantum fluctuations of the scalar field) leading eventually to the
large scale structure formation in the Universe. It is also
predicted that in a simple slow-roll inflation (defined by a
canonical scalar field with a nearly flat potential) the dominant
mode of the primordial density perturbations is almost adiabatic and
nearly scale invariant with a Gaussian
distribution~\cite{Lin90,Lid00a,Lid97,Rio02,Lyt09,Mal03}.

A problem with the inflation models is that the width of the
potential must be much larger than its height, so that there will be a large number of e-folds of the scale factor to
fit with the CMB anisotropy measurements. In fact, from Ref.~\cite{Ada91} the ratio between the height and the fourth power of the width must fulfill
\begin{equation}
\label{eq1} \frac{\Delta V}{(\Delta\phi)^{4}}\leq 10^{-6}
\end{equation}
which means that the potential is almost flat. In this relation
$\Delta$ shows the change in the corresponding parameters. In this
regard, to address the theoretical problems of the rolling inflaton
models, Freese, Frieman, and Olinto in 1990 have proposed the
Natural Inflation (NI) scenario~\cite{Fre90}. In the Natural
Inflation, an axion-like particle (a pseudo-Nambu-Goldstone boson)
is the field responsible for running of the inflation. A shift
symmetry (that is, the potential is invariant under a transformation
$\phi\rightarrow \phi+constant$) in the Natural Inflation ensures
flatness of the potential~\cite{Fre90,Fre04,Fre14}. Note that in
this model the potential is ``nearly'' flat and eventually, after
enough inflation, the symmetry is broken and the inflation phase
terminates. Natural Inflation, in its simplest realization, has the
following potential
\begin{equation}
\label{eq2}
V(\phi)=\Lambda^{4}\bigg[1+\cos\bigg(\frac{\phi}{f}\bigg)\bigg].
\end{equation}
The height of the potential is given by $2\Lambda^{4}$ and the width
by $\pi f$. In fact, $f$ is an axion decay constant
which parameterizes the spontaneous symmetry breaking scale needed
to end the inflation phase. Actually, when a global symmetry is
spontaneously broken in particle physics, Nambu-Goldstone bosons
arise at a scale $f\sim m_{pl}$ (where $m_{pl}$ is the reduced Planck mass). However, when the shift symmetry is
exact, the inflaton doesn't roll and therefore inflation doesn't
happen. By an explicit symmetry breaking, the pseudo-Nambu-Goldstone
bosons with ``nearly flat'' potential arise and inflation can
happen. These bosons arise at a mass scale $m\sim \Lambda$ below the spontaneous symmetry breaking
scale~\cite{Ada92}. These two scales are important because they show
the symmetry breaking's scales. To constraint the values of
$\Lambda$ and $f$, one can use the recent bounds on the scalar
spectral index and tensor-to-scalar ratio from observational data.

In the original Natural Inflation (a canonical scalar field with
\emph{cosine} potential) and with $\Lambda\sim m_{GUT}\sim10^{16}$
GeV, the pseudo-Nambu-Goldstone boson field runs inflation if $f\geq
m_{pl}$. In this limit, the scalar spectral index and
tensor-to-scalar ratio of the Natural Inflation are consistent with
observational data. For $f\sim m_{pl}$ and considering $m_{pl}\sim
10^{19}$ GeV, we have the height-to-forth power of the width ratio
as $\frac{\Lambda^{4}}{f^{4}}\sim 10^{-12}$ (which satisfies the
condition (\ref{eq1})). In the $f\gg m_{pl}$ limit, $n_{s}$ and $r$
are independent of $f$ and for a given value of the number of e-folds
there is a specific fixed point in $r-n_{s}$ plane. Actually, in
this limit, Natural Inflation meets a large field $m^{2}\phi^{2}$
model. In this case, one gets $m_{\phi}\sim10^{13}$
GeV~\cite{Fre14,Fre93}. The authors of Ref.~\cite{Co15}, by
considering the new bounds on the scalar spectral index from
Planck2015 and by studying the reheating phase with Natural potential,
have obtained a new constraint on $f$ as $f>5.6\, m_{pl}$. The authors
of Ref.~\cite{Ger17} have studied the Hybrid Natural Inflation and
by considering the energy scale of the inflation they have obtained
some bounds on $f$ in some cases. See also~\cite{Ger10,Vis11,Mis11}
for other works on the Natural Inflation.

The mentioned works have considered a canonical scalar field with
cosine potential. However, it is believed that there is a
possibility that inflation may be driven by a single field where its
kinetic energy is non-canonical. Such non-canonical models (usually
referred as ``k-inflation'') predict that the primordial density
perturbations are somehow scale dependent (mildly supported by the
Planck2015 observational data~\cite{Ade15a,Ade15b}) and their
distribution is non-Gaussian. Among the k-inflation models, we can
mention the Tachyon inflation where the tachyon field is associated with the
D-branes in string theory~\cite{Sen99,Sen02a,Sen02b}. When this
field rolls slowly down its potential, the Universe evolves smoothly
from an accelerating expansion phase to a nonrelativistic fluid
dominated era~\cite{Gib02}. The tachyon, as a dark energy
component, can be responsible for the late time cosmic speed-up of
the Universe~\cite{Pad02,Gor04,Cop05}. It can also be considered as
the inflaton driving the initial cosmological inflation
phase~\cite{Sam02,Fei02}. Some aspects of tachyon field cosmology
can be seen in~\cite{Noz13a,Noz14}.

On the other hand, it has been shown that in a moving
\textit{Dp}-brane in the $k$ \textit{NS5}-brane background (around a
ring with radius $R$, where $k$ is the number of \textit{NS5}-branes) the radion could be tachyonic. This is the
idea of the geometrical tachyon~\cite{Kut04,Tho05a,Tho05b,Tho05c}.
If one considers the solution inside the ring and uses a tachyon map,
the potential of the tachyon would be of the cosine type as $V(\phi)=A
\cos\left(\frac{\phi}{kl_{s}^2}\right)$ with $A=\frac{\tau
	R}{\sqrt{kl_{s}^2}}$~\cite{Tho05b,Tho05c}. Here $\tau$ is the tension of
the \textit{Dp}-brane, $R$ the radius of the ring and  $l_{s}$
the string length. By comparing this potential with the potential in
Natural Inflation we see that the string length is related to the
width of the potential, and tension and radius are related to its
height. Roughly speaking, we can take $kl_{s}^2\equiv f$ and
$A\equiv \Lambda^{4}$.

Here we consider a tachyon field with a Natural potential as driver
of cosmological inflation. Adopting the potential of the Natural
Inflation is just like a constant shift in the ``cosine'' potential
obtained by considering a probe \textit{Dp}-brane in a
\textit{NS5}-brane ring background. Actually, one of the problems of
the open string tachyon is that its potential should tend to zero as
$\phi\rightarrow \infty$ in order that no \textit{D}-brane and open
string should exist at ground state~\cite{Kof02,Fro02}. So, in this case there
will be no reheating phase because the potential has a minimum at
asymptotic infinity. However, by adopting the potential
(\ref{eq2}), we shall not encounter with this issue in the sense
that in this case there would be a minimum at a finite value of
$\phi$. As we shall see, Tachyon Natural Inflation (TNI) in the large
$f$ limit approaches the tachyon model with $\phi^{2}$ potential which
is consistent with the Planck2015 observational data. Note that the
Planck2015 observational data rule out a canonical scalar field
with $\phi^{2}$ potential. Actually, although the scalar spectral
index of the canonical $\phi^{2}$ model is consistent with
observation, its tensor-to-scalar ratio is out of the $95\%$ CL of the
Planck2015 observational data. However, as we show, the tachyon
$\phi^{2}$ model is consistent with the Planck2015 dataset. We study
cosmological inflation and perturbations in a TNI model with warped
background geometry (specified by the warp factor $\lambda$). By a
numerical study of the scalar spectral index and the
tensor-to-scalar ratio we obtain some new constraints on parameter
$f$. The reason that we consider a warped background is that to have
tachyon inflation with ``steep potential" it is necessary that the
background is warped~\cite{Che05}. We shall see that the bound on
$f$ is not just in terms of $m_{pl}$, as it is in the canonical NI.
In the TNI model, the constraints on $f$ depends on $m_{pl}$,
$\lambda$ and $\Lambda$. This is because of the form of the energy
density (and Friedmann equation) in the tachyon model.

Exploring the reheating process after the end of inflation is an
important subject in studying the cosmic inflation. After the
Universe inflates sufficiently and the slow-roll conditions break
down, the inflation era terminates. By ending the inflation phase,
the scalar field responsible for cosmic inflation starts to
oscillate about the minimum of its potential. The simple canonical
reheating scenario states that the inflaton loses its energy by
oscillation and decays into a plasma of the relativistic particles
(corresponding to a radiation dominated Universe) by entering the
processes which include the physics of non-equilibrium phenomena and
particle creation~\cite{Ab82,Do82,Al82}. However, some other
complicated scenarios of reheating including non-perturbative
processes have been proposed by several authors. The tachyonic
instability~\cite{Gr97,Sh06,Du06,Ab10,Fel01a,Fel01b}, the instant
preheating~\cite{Fel99} and the parametric resonance
decay~\cite{Ko94,Tr90,Ko97} are some examples of the
non-perturbative reheating scenarios which should be mentioned. One can
characterize the reheating era dynamics by seeking for the reheating
temperature ($T_{rh}$) and the number of e-folds during reheating
($N_{rh}$) which give some more constraints on the model
parameters~\cite{Dai14,Un15,Co15,Cai15,Ue16,Noz16b}. The effective
equation of state parameter during reheating ($\omega_{eff}$) is
another important parameter which gives us some more useful
information. Domination of the potential energy of the field over the
kinetic energy gives $\omega_{eff}=-1$ and domination of the kinetic
term over potential energy gives $\omega_{eff}=+1$. We assume the
range of the effective equation of state parameter during the
reheating phase to be given as
$-\frac{1}{3}\leq\omega_{eff}\leq\frac{1}{3}$. This is because, we
have $\omega_{eff}=-\frac{1}{3}$ at the end of the inflation era and
$\omega_{eff}=\frac{1}{3}$ at the beginning of the radiation
dominated era. At the initial stage of the reheating era, the
oscillation frequency of the massive inflaton is much larger than
the expansion rate. This situation leads to a vanishing averaged
effective pressure which can effectively be considered as the
equation of state parameter of the matter. Then, by oscillating and
decaying the inflaton field into other particles, there would be an
increment in the value of $\omega_{eff}$ with time. The effective
equation of state parameter increases until at the beginning of the
radiation domination era, when it reaches $\frac{1}{3}$. Seeking for
the effective equation of state parameter helps us to obtain some
more constraints on the model's parameter space. Ref.~\cite{Am15} is
a seminal review article on the reheating issue.

With these preliminaries, this paper is devoted to an extension of
Natural Inflation in the spirit of models of inflation with non-canonical scalar fields.
We assume that the cosmological inflation is driven
by a tachyon field in the framework of Natural Inflation, and
investigate the viability and observational status of this model by
focusing on the primordial perturbations and also reheating in this
framework. This paper is organized as follows: In section 2 we
consider a tachyon model with \emph{cosine} potential and study the
inflation and perturbation in this setup. We obtain the main
perturbation parameters such as the scalar spectral index, its
running and the tensor-to-scalar ratio and study these parameters
numerically. By comparing the numerical results with both $68\%$ and
$95\%$ CL regions of the Planck2015 TT, TE, EE+lowP data, we obtain
some constraints on the width of the potential ($f$). In section 3
we study the reheating in this Tachyonic Natural Inflation. We
obtain some expressions for e-folds number and temperature in this
era in terms of the scalar spectral index and the effective equation
of state parameter. By considering the values of the effective
equation of state parameter, we obtain some constraints on the width
of the potential as well as the tensor-to-scalar ratio. Also in this
section, we obtain some constraints on the number of e-folds and
temperature during reheating based on the observationally viable
values of the scalar spectral index. In section 4 we present a
summary and conclusions.

\section{Inflation}
The action for a tachyon inflation model is given by the following expression
\begin{eqnarray}
\label{eq3} S=\int
d^{4}x\sqrt{-g}\Bigg[\frac{m_{pl}^{2}}{2}R-V(\phi)\sqrt{1-2\lambda\,
	X} \Bigg],
\end{eqnarray}
where $R$ is the Ricci scalar, $m_{pl}$ is the reduced Planck mass,
$\lambda$ is the constant warp factor and $X=-\frac{1}{2}g^{\mu\nu}\partial_{\mu}\phi\,\partial_{\nu}\phi$\,. We assume the potential of
the tachyon scalar field to be as given in equation (\ref{eq2}). In a
spatially flat FRW metric, by using action (\ref{eq3}) we get the
Friedmann equation of the model as follows
\begin{equation}
\label{eq4}H^{2}=\frac{1}{3m_{pl}^{2}}\frac{V}{\sqrt{1-\lambda\dot{\phi}^{2}}}\,,
\end{equation}
where a cosmic time derivative is denoted by a dot. By varying the
action (\ref{eq3}) with respect to the tachyon field, we derive the
following equation of motion
\begin{equation}
\label{eq5}\frac{\ddot{\phi}}{1-\dot{\phi}^{2}}+3H\dot{\phi}+\frac{V'}{\lambda
	V}=0\,,
\end{equation}
where a derivative with respect to the tachyon field is shown by a
prime. Satisfying the slow-roll conditions $\epsilon \ll 1$ and
$\eta \ll 1$, where
\begin{equation}
\label{eq6}\epsilon\equiv-\frac{\dot{H}}{H^{2}} \,,
\end{equation}
and
\begin{equation}
\label{eq7}\eta\equiv-\frac{1}{H}\frac{\ddot{H}}{\dot{H}}\,,
\end{equation}
are the slow-roll parameters, must be satisfied in order to have an inflation phase. These parameters are much
smaller than unity in the inflationary era and the inflation ends
when one of them reaches the unity. The number of e-folds during
inflation is given by
\begin{equation}
\label{eq8} N=\int_{t_{hc}}^{t_{end}} H dt
\end{equation}
where the subscripts $hc$ and $end$ mark the time of the horizon
crossing and end of inflation respectively. Equations
(\ref{eq4})-(\ref{eq8}) are the background equations of the model.

Now to obtain the perturbation parameters, we use the following ADM
perturbed metric
\begin{eqnarray}
\label{eq9} ds^{2}=
-(1+2{\Phi})dt^{2}+2a(t){\cal{Y}}_{i}\,dt\,dx^{i}
+a^{2}(t)\left[(1-2{{\cal{Z}}})\delta_{ij}+2{\Theta}_{ij}\right]dx^{i}dx^{j}\,.
\end{eqnarray}
In this relation
${\cal{Y}}^{i}=\delta^{ij}\partial_{j}{\cal{Y}}+v^{i}$ where $v^{i}$
is a vector which satisfies the condition $v^{i}_{,i}=0 $. ${\Phi}$
and ${\cal{Y}}$ are 3-scalars. ${\Theta}_{ij}$ is defined as a
spatial symmetric and traceless shear 3-tensor and ${{\cal{Z}}}$ is
the spatial curvature perturbation. The uniform-field gauge
(characterized by $\delta\phi=0$) is a convenient gauge to study the
scalar perturbation of the theory. By working within this gauge and
assuming ${\Theta}_{ij}=0$, we get~\cite{Muk92,Bau09,Bar80}
\begin{eqnarray}
\label{eq10}
ds^{2}=-(1+2{\Phi})dt^{2}+2a(t){\cal{Y}}_{,i}\,dt\,dx^{i}
+a^{2}(t)(1-2{{\cal{Z}}})\delta_{ij}dx^{i}dx^{j}\,,
\end{eqnarray}
where we have considered the scalar part of the perturbation. By
using this metric, we can expand the action up to the second order
in perturbations as
\begin{eqnarray}
\label{eq11} S_{2}=\int
dt\,d^{3}x\,a^{3}\,{\cal{W}}_{s}\Bigg[\dot{\cal{Z}}-\frac{c_{s}^2}{a^{2}}(\partial
{\cal{Z}})^{2}\Bigg]\,,
\end{eqnarray}
where
\begin{equation}
\label{eq12} {\cal{W}}_{s}=\frac {4\lambda\,\dot{\phi}^{2}\,V}{
	\big( 1-4\lambda\,\dot{\phi}^{2}\big) ^{\frac{3}{2}}{H}^{2}}\,,
\end{equation}
and
\begin{equation}
\label{eq13} c_{s}=\sqrt{1-4\lambda\,\dot{\phi}^{2}}\,.
\end{equation}
The parameter $c_{s}$ is the sound speed. The two-point correlation
function, which helps us to survey the power spectrum, is given by
\begin{equation}
\label{eq14} \langle
0|{{\cal{Z}}}(0,\textbf{k}_{1}){{\cal{Z}}}(0,\textbf{k}_{2})|0\rangle
=(2\pi)^{3}\delta^{3}(\textbf{k}_{1}+\textbf{k}_{2})\frac{2\pi^{2}}{k^{3}}{\cal{A}}_{s}\,.
\end{equation}
In this relation, the parameter ${\cal{A}}_{s}$ is the power
spectrum which is defined as follows (see
Refs.~\cite{Noz13a,Fel11a,Fel11b,Che08} for details)
\begin{equation}
\label{eq15}{\cal{A}}_{s}=\frac{H^{2}}{8\pi^{2}{\cal{W}}_{s}c_{s}^{3}}\,.
\end{equation}
This power spectrum leads to the following scalar spectral index
\begin{eqnarray}
\label{eq16} n_{s}-1=\frac{d \ln {\cal{A}}_{s}}{d \ln
	k}\Bigg|_{c_{s}k=aH}=-6\epsilon+2\eta-s\,,
\end{eqnarray}
where, $k$ is the wave number of the perturbation, and
\begin{equation}\label{eq17}
s=\frac{1}{H}\frac{d \ln c_{s}}{dt}\,.
\end{equation}
In our TNI model, the running of the scalar spectral index is given
by
\begin{eqnarray}
\label{eq18} \alpha_{s}=\frac{d \ln n_{s}}{d \ln
	k}\Bigg|_{c_{s}k=aH}=-2\lambda^{2}\,\zeta+24\lambda^{2}\,\epsilon\,\eta-24\lambda^{2}\epsilon^{2}
+m_{pl}^{2}\frac{\lambda V'}{V^{2}}s'\,,
\end{eqnarray}
where
\begin{equation}
\label{eq19} \zeta=m_{pl}^{4}\,\frac{V'\,V'''}{\lambda^{2}V^{4}}\,.
\end{equation}

To study the tensor part of the theory, we focus on the tensor part
of the perturbed metric (\ref{eq11}) and write the 3-tensor
${\Theta}_{ij}$ as
\begin{equation}
\label{eq20}
{\Theta}_{ij}={\Theta}_{+}\varepsilon_{ij}^{+}+{\Theta}_{\times}\varepsilon_{ij}^{\times}\,.
\end{equation}
The two polarization tensors in the above relation
($\varepsilon_{ij}^{+}$ and $\varepsilon_{ij}^{\times}$) satisfy the
reality and normalization conditions~\cite{Fel11a,Fel11b}. We obtain
the quadratic action for the tensor mode of the perturbations
(gravitational waves) as follows
\begin{eqnarray}
\label{eq21} S_{T}=\int dt\, d^{3}x\, a^{3}
{\cal{W}}_{T}\left[\dot{\Theta}_{+}^{2}-\frac{(\partial
	{\Theta}_{+})^{2}}{a^{2}}+\dot{\Theta}_{\times}^{2}-\frac{(\partial
	{\Theta}_{\times})^{2}}{a^{2}}\right]\,,
\end{eqnarray}
where ${\cal{W}}_{T}=m_{pl}^{4}$. In this regard, similar to the
scalar part, we have the amplitude of the tensor perturbations as
\begin{equation}
\label{eq22} {\cal{A}}_{T}=\frac{H^{2}}{2\pi^{2}{\cal{W}}_{T}}\,,
\end{equation}
which gives the tensor spectral index in our TNI model as follows
\begin{equation}
\label{eq23} n_{T}=\frac{d \ln {\cal{A}}_{T}}{d \ln k}=-2\epsilon\,.
\end{equation}
We note that to measure the tensor spectral index, a detection of
the CMB B-mode polarization is required. The accuracy of the current
experiments is not enough to detect this mode in observation. In
Refs.~\cite{Sim15,Boy15} the authors have forecasted future CMB
polarization experiments that would be able to measure the tensor
spectral index in essence.

Another important perturbation parameter is the tensor-to-scalar
ratio which in this TNI setup is given by
\begin{eqnarray}
\label{eq24} r=16c_{s}\epsilon\,.
\end{eqnarray}

To investigate the cosmological viability of the model and also to
find some constraints on the model's parameter space, we treat the
perturbation parameters numerically. In this regard, and to obtain
the values of these perturbation parameters at horizon crossing, we
should use the equation (\ref{eq8}). To this end, we first find the
value of the scalar field at the end of the inflation by setting
$\epsilon=1$. From now on, we work within the slow-roll limit
($\ddot{\phi}\ll 3H\dot{\phi}$ and $\lambda\dot{\phi}^{2}\ll 1$).
With this approximation, from equations (\ref{eq6}) and (\ref{eq7}),
we have
\begin{equation}
\label{eq25}\epsilon=\frac{m_{pl}^{2}}{2\lambda}\frac{V'^{2}}{V^{3}}=-\frac{1}{2\beta}\,
\frac{\cos \left( {\frac {\phi}{f}} \right) -1}{ \left( \cos \left(
	{\frac { \phi}{f}} \right)  \right) ^{2}+2\,\cos \left( {\frac
		{\phi}{f}} \right) +1} \,,
\end{equation}
where
\begin{equation}
\label{eq26}\beta=\Lambda^{4}f^{2}\lambda m_{pl}^{-2}\,,
\end{equation}
and
\begin{equation}
\label{eq27}\eta=\frac{m_{pl}^{2}}{\lambda}\left[\frac{V''}{V^{2}}-\frac{1}{2}\frac{V'^{2}}{V^{3}}\right]=-\frac{1}{2\beta}\,
\left[ 1+\cos \left( {\frac {\phi}{f}} \right) \right] ^{-1}\,.
\end{equation}
By considering the potential given by equation (\ref{eq2}), the
number of e-folds in TNI model and with the slow-roll approximation,
takes the following form
\begin{eqnarray}
\label{eq28}N=\beta\Bigg[\cos \left( {\frac {\phi_{end}}{f}} \right)
-\cos \left( {\frac {\phi_{hc}}{f}} \right)\Bigg] +2\beta\,\ln
\left[\frac{ \cos \left( {\frac {\phi_{end}}{f}} \right) -1 }{ \cos
	\left( {\frac {\phi_{hc}}{f }} \right) -1} \right]\,.
\end{eqnarray}
By setting $\epsilon=1$, we get
\begin{equation}
\label{eq29}\cos
\bigg(\frac{\phi_{end}}{f}\bigg)=\frac{1}{4}\,{\frac
	{-4\,\beta-1+\sqrt {16\,\beta+1}}{\beta}}.
\end{equation}
Now, by using equations (\ref{eq28}) and (\ref{eq29}) we find
\begin{equation}
\label{eq30}\cos \bigg(\frac{\phi_{hc}}{f}\bigg)=1+{\cal{G}}\,,
\end{equation}
where
\begin{eqnarray}
\label{eq31}{\cal{G}}=2\,{\it LambertW} \Bigg( \frac{1}{8}\, \left(
-1+\sqrt {16\,\beta+1}-8\,\beta \right)  {{\rm
		e}^{-\frac{1}{8}\,{\frac {8\,\beta+4\,N-\sqrt
				{16\,\beta+1}+1}{\beta}}}}{\beta}^{-1} \Bigg)\,.
\end{eqnarray}
Equation (\ref{eq30}) implies the constraint $-2\leq{\cal{G}}\leq
0$. In the slow-roll limit, we have $c_{s}^{2}=1$ and so $s=s'=0$.
In this regard, by substituting equation (\ref{eq30}) into equations
(\ref{eq16}), (\ref{eq18}) and (\ref{eq24}), we find the
perturbation parameters in the slow-roll limit as (see for instance, \cite{Gron})
\begin{equation}
\label{eq32}n_{s}=1+\frac{2}{\beta}\frac{{\cal{G}}-1}{({\cal{G}}+2)^{2}}\,,
\end{equation}

\begin{equation}
\label{eq33}r=-\frac{8}{\beta}\frac{{\cal{G}}}{({\cal{G}}+2)^{2}}\,,
\end{equation}
\begin{equation}
\label{eq34}\alpha_{s}=\,\frac{2}{\beta^{2}}\frac{(4-{\cal{G}}){\cal{G}}}{({\cal{G}}+2)^{4}}\,.
\end{equation}
which are written in terms of the parameter $\beta$.

\begin{figure}[ht]
\begin{center}
\includegraphics[scale=0.6]{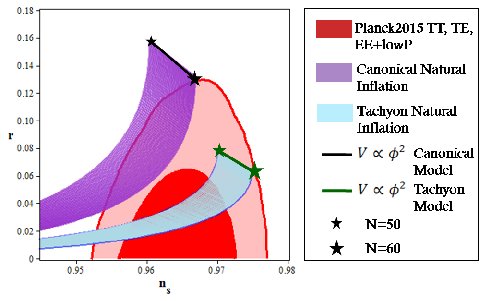}
\end{center}
\caption{\small {The tensor-to-scalar ratio versus the
scalar spectral index in the background of the Planck2015 TT, TE,
EE+lowP data.}}
\label{fig1}
\end{figure}

\begin{figure}[ht]
\begin{center}
\includegraphics[scale=0.65]{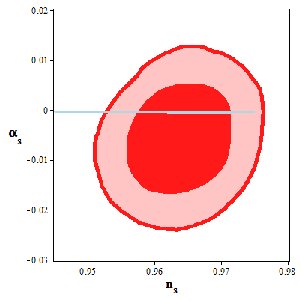}
\includegraphics[scale=0.65]{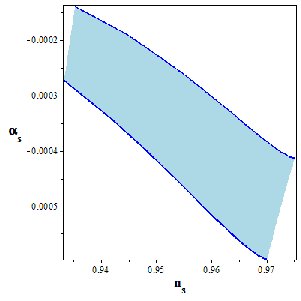}
\end{center}
\caption{\small {Running of the scalar spectral index
versus the scalar spectral index in the background of the Planck2015
TT, TE, EE+lowP data. In the right panel we have zoomed out the cyan
region of the left panel.}}
\label{fig2}
\end{figure}

After obtaining these quantities, we can perform a numerical
analysis on the model's parameter space. To predict the values of
$n_{s}$, $r$ and $\alpha_{s}$, we adopt $50\leq N\leq 60$. By
plotting $r-n_{s}$ and $\alpha_{s}-n_{s}$ planes in the background
of $68\%$ CL and $95\%$ CL regions of the Planck2015 TT, TE, EE+lowP
data, we obtain some constraints on the parameter $\beta$. In figure
1 the region with violet color corresponds to the canonical Natural
Inflation and the cyan region corresponds to our Tachyon Natural
Inflation. We see that the Tachyon Natural Inflation is more
consistent with observational data than the canonical Natural Inflation.
The left panel of figure 2 shows the evolution of the running of the
scalar spectral index with respect to the scalar spectral index in
the background of the Planck2015 TT, TE, EE+lowP data. In the right
panel of figure 2 we have zoomed out the cyan region of the left
panel. Our analysis shows that for $N=50$, the parameters $n_{s}$, $\alpha_{s}$ and $r$ are consistent
with the Planck2015 data if $f>
\sqrt{\textbf{13}\lambda^{-1}}\,\Lambda^{-2}\,m_{pl}$. For $N=60$,
these parameters are consistent with the Planck2015 data if $f>
\sqrt{\textbf{12}\lambda^{-1}}\,\Lambda^{-2}\,m_{pl}$. We see that
the bound on $f$ in TNI is in terms of $m_{pl}$, $\Lambda$ and
$\lambda$ (contrary to the canonic NI in which the constraint on
$f$ is only in terms of $m_{pl}$). Note that, if $\lambda\sim
\Lambda^{-2}$, we get $\frac{f^{4}}{\Lambda^{4}}>
(\sqrt{\textbf{13}}\,m_{pl})^{4}$ for $N=50$ and
$\frac{f^{4}}{\Lambda^{4}}> (\sqrt{\textbf{12}}\,m_{pl})^{4}$ for
$N=60$. In this respect, these constraints satisfy the condition
(\ref{eq1}). Table 1 gives a summary of constraints imposed on $f$
in our setup based on the observationally viable values of the
scalar spectral index, running of the scalar spectral index and the
tensor-to-scalar ratio for both $68\%$ and $95\%$ CL of the
Planck2015 TT, TE, EE+lowP data.  Table 2 is the same as table 1,
but now by setting $\lambda\sim \Lambda^{-4}$ in order to state the
obtained constraints just in terms of the reduced Planck mass.

\begin{table*}
\caption{\label{tab1} Constraints on $f$ based on the
observationally viable values of the scalar spectral index, its
running and the tensor-to-scalar ratio.}
\begin{tabular}{cccccccc}
\\ \hline \hline&$N=50$&&$N=55$
&&$N=60$&&\\ \hline\\
$r-n_{s}\,,\,68\%$ CL&
$\sqrt{\frac{18}{\lambda}}\frac{m_{pl}}{\Lambda^{2}}\leq f\leq
\sqrt{\frac{90}{\lambda}}\frac{m_{pl}}{\Lambda^{2}}$
&&$\sqrt{\frac{15}{\lambda}}\frac{m_{pl}}{\Lambda^{2}}\leq f\leq
\sqrt{\frac{65}{\lambda}}\frac{m_{pl}}{\Lambda^{2}}$&&$\sqrt{\frac{14}{\lambda}}\frac{m_{pl}}{\Lambda^{2}}\leq
f\leq \sqrt{\frac{40}{\lambda}}\frac{m_{pl}}{\Lambda^{2}}$
&&\\\\
$r-n_{s}\,,\,95\%$ CL&
$\sqrt{\frac{13}{\lambda}}\frac{m_{pl}}{\Lambda^{2}}\leq f$
&&$\sqrt{\frac{12.5}{\lambda}}\frac{m_{pl}}{\Lambda^{2}}<
f$&&$\sqrt{\frac{12}{\lambda}}\frac{m_{pl}}{\Lambda^{2}}\leq f$
&&\\\\
\hline\\ $\alpha_{s}-n_{s}\,,\,68\%$ CL&
$\sqrt{\frac{17}{\lambda}}\frac{m_{pl}}{\Lambda^{2}}\leq f$
&&$\sqrt{\frac{15}{\lambda}}\frac{m_{pl}}{\Lambda^{2}}\leq f\leq
\sqrt{\frac{75}{\lambda}}\frac{m_{pl}}{\Lambda^{2}}$&&$\sqrt{\frac{14}{\lambda}}\frac{m_{pl}}{\Lambda^{2}}\leq
f\leq \sqrt{\frac{45}{\lambda}}\frac{m_{pl}}{\Lambda^{2}}$
&&\\\\
$\alpha_{s}-n_{s}\,,\,95\%$ CL&
$\sqrt{\frac{13}{\lambda}}\frac{m_{pl}}{\Lambda^{2}}\leq f$
&&$\sqrt{\frac{12.5}{\lambda}}\frac{m_{pl}}{\Lambda^{2}}<
f$&&$\sqrt{\frac{12}{\lambda}}\frac{m_{pl}}{\Lambda^{2}}\leq f$
&&\\\\
\hline
\end{tabular}
\end{table*}

\begin{table*}
\caption{\label{tab2} Constraints on $f$ based on the
observationally viable values of the scalar spectral index, its
running and the tensor-to-scalar ratio by adopting
$\lambda=\Lambda^{-4}$.}
\begin{tabular}{cccccccc}
\\ \hline \hline&$N=50$&&$N=55$
&&$N=60$&&\\ \hline\\
$r-n_{s}\,,\,68\%$ CL& $\sqrt{18}\,m_{pl}\leq f\leq
\sqrt{90}\,m_{pl}$ &&$\sqrt{15}\,m_{pl}\leq f\leq
\sqrt{65}\,m_{pl}$&&$\sqrt{14}\,m_{pl}\leq f\leq \sqrt{40}\,m_{pl}$
&&\\\\
$r-n_{s}\,,\,95\%$ CL& $\sqrt{13}\,m_{pl}\leq f$
&&$\sqrt{12.5}\,m_{pl}< f$&&$\sqrt{12}\,m_{pl}\leq f$
&&\\\\
\hline\\ $\alpha_{s}-n_{s}\,,\,68\%$ CL& $\sqrt{17}\,m_{pl}\leq f$
&&$\sqrt{15}\,m_{pl}\leq f\leq
\sqrt{75}\,m_{pl}$&&$\sqrt{14}\,m_{pl}\leq f\leq \sqrt{45}\,m_{pl}$
&&\\\\
$\alpha_{s}-n_{s}\,,\,95\%$ CL& $\sqrt{13}\,m_{pl}\leq f$
&&$\sqrt{12.5}\,m_{pl}< f$&&$\sqrt{12}\,m_{pl}\leq f$
&&\\\\
\hline
\end{tabular}
\end{table*}

\begin{figure}[ht]
\begin{center}
\includegraphics[scale=0.4]{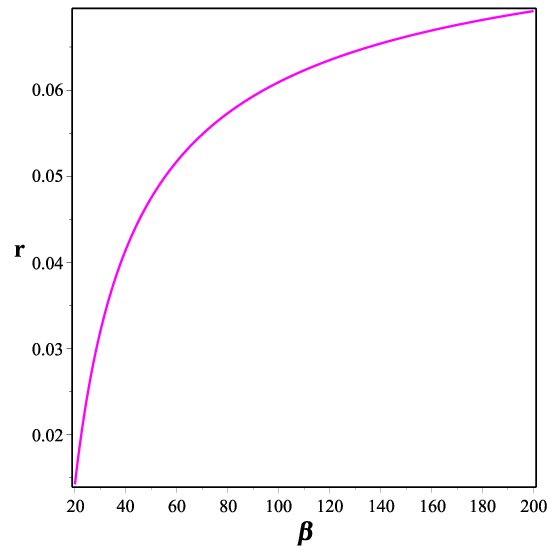}
\end{center}
\caption{\small {Tensor-to-scalar ratio versus the
parameter $\beta$, for $\delta_{ns}=0.032$.}}
\label{fig3}
\end{figure}

We can also find a constraint on the tensor-to-scalar ratio in terms
of the scalar spectral tilt, $\delta_{ns}=1-n_{s}$. By solving
equation (\ref{eq32}) for ${\cal{G}}$, we find
\begin{equation}\label{eq35}
{\cal{G}}=-\frac{1+2\beta\delta_{ns}-\sqrt{1+6\beta\,\delta_{ns}}}{\beta\,\delta_{ns}}\,,
\end{equation}
where the minus sign has been chosen due to the condition
$-2\leq{\cal{G}}\leq 0$, which leads to the requirement
$\beta>\frac{1}{2\delta_{ns}}$ or $
f\geq\sqrt{\frac{1}{2\lambda\delta_{ns}}}\frac{m_{pl}}{\Lambda^{2}}$.
With $n_{s}=0.968$, resulting in $\delta_{ns}=0.032$, the
requirement $\beta\geq\frac{1}{2\delta_{ns}}$ demands the constraint
$\beta\geq15.6$ or
$f\geq\sqrt{\frac{15.6}{\lambda}}\frac{m_{pl}}{\Lambda^{2}}$.
Inserting expression (\ref{eq35}) into equation (\ref{eq33}) gives
\begin{equation}\label{eq36}
r=8\,\delta_{ns}\,\frac{1+2\beta\delta_{ns}-\sqrt{1+6\beta\,\delta_{ns}}}{\left(1-\sqrt{1+6\beta\,\delta_{ns}}\right)^{2}}\,.
\end{equation}
In figure 3 we have plotted the tensor-to-scalar ratio versus the
parameter $\beta$, for $\delta_{ns}=0.032$. As figure shows, $r$ is
an increasing function of $\beta$, with
\begin{equation}\label{eq37}
\lim_{\beta\rightarrow\infty}r=\frac{8}{3}\,\delta_{ns}\,.
\end{equation}
Therefore, a prediction of this model is
$r\leq\frac{8}{3}\,\delta_{ns}$. This prediction becomes
$r\leq0.085$ for $\delta_{ns}=0.032$.

\section{Reheating}

One important stage in inflation models is the reheating at the end
of inflation. It is necessary to warm up the universe sufficiently
for subsequent processes. In this regard, we study this process in
our TNI model and obtain some more constraints on the model's
parameter space. From the strategy used in
Refs.~\cite{Dai14,Un15,Co15,Cai15,Ue16}, we can express the
reheating parameters $N_{rh}$ and $T_{rh}$ in terms of the scalar
spectral index in the TNI model. The relation between the number of e-folds
and the scale factor at the horizon crossing ($a_{hc}$) as well as
the scale factor at the end of inflation ($a_{end}$) is given by
\begin{equation}
\label{eq38} N=\ln \left(\frac{a_{end}}{a_{hc}}\right)\,.
\end{equation}
By introducing $\omega_{eff}$ as the effective equation of state
parameter of the dominant component of the cosmic energy during the reheating epoch, we
have $\rho\sim a^{-3(1+\omega_{eff})}$ for the energy density. Now,
by using the scale factors at the end of inflation and the reheating
era, we can write the number of e-folds during the reheating in terms
of the energy density and the effective equation of state parameter
in this era as follows
\begin{eqnarray}\label{eq39}
N_{rh}=\ln\left(\frac{a_{rh}}{a_{end}}\right)=-\frac{1}{3(1+\omega_{eff})}\ln\left(\frac{\rho_{rh}}{\rho_{end}}\right)\,.
\end{eqnarray}
By assuming $k_{hc}$ to be the value of $k$ at the horizon crossing
and $a_{0}$ as the current value of the scale factor, we get
\begin{eqnarray}\label{eq40}
0=\ln\left(\frac{k_{hc}}{a_{hc}H_{hc}}\right)=
\ln\left(\frac{a_{end}}{a_{hc}}\frac{a_{rh}}{a_{end}}\frac{a_{0}}{a_{rh}}\frac{k_{hc}}{a_{0}H_{hc}}\right)\,.
\end{eqnarray}
The following expression is obtained from equations (\ref{eq38}),
(\ref{eq39}) and (\ref{eq40})
\begin{eqnarray}\label{eq41}
N+N_{rh}+\ln\left(\frac{k_{hc}}{a_{0}H_{hc}}\right)+\ln\left(\frac{a_{0}}{a_{rh}}\right)=0\,.
\end{eqnarray}
We can express the term $\frac{a_{0}}{a_{rh}}$ in terms of the
temperature and energy density in the reheating era. To this end, we
use the relation between energy density and temperature in reheating
era as~\cite{Co15,Ue16}
\begin{equation}\label{eq42}
\rho_{rh}=\frac{\pi^{2}g_{rh}}{30}T_{rh}^{4}\,,
\end{equation}
with $g_{rh}$ to be the effective number of the relativistic species
at the reheating era. Also, we can relate the scale factor at the
reheating era to the temperature in this era by using the following
equation~\cite{Co15,Ue16}
\begin{equation}\label{eq43}
\frac{a_{0}}{a_{rh}}=\left(\frac{43}{11g_{rh}}\right)^{-\frac{1}{3}}\frac{T_{rh}}{T_{0}}\,.
\end{equation}
In this relation the current temperature of the Universe is shown by
$T_{0}$. By using equation (\ref{eq42}) and (\ref{eq43}) we find
\begin{eqnarray}\label{eq44}
\frac{a_{0}}{a_{rh}}=\left(\frac{43}{11g_{rh}}\right)^{-\frac{1}{3}}T_{0}^{-1}\left(\frac{\pi^{2}g_{rh}}{30\rho_{rh}}\right)^{-\frac{1}{4}}\,.
\end{eqnarray}
The energy density in our TNI model can be written in terms of the
slow-roll parameter as
\begin{eqnarray}\label{eq45}
\rho=\frac{V}{\sqrt{1-\frac{2}{3\lambda}\epsilon}}\,.
\end{eqnarray}
Setting $\epsilon=1$ gives the energy density at the end of
inflation phase as follows
\begin{equation}\label{eq46}
\rho_{end}=\sqrt{\frac{3\lambda}{3\lambda-2}}\,V_{end}=\sqrt{\frac{3\lambda}{3\lambda-2}}\,\Lambda^{4}\Bigg[1
+\cos\bigg(\frac{\phi_{end}}{f}\bigg)\Bigg]\,.
\end{equation}
By using equations (\ref{eq39}) and (\ref{eq46}) we get
\begin{eqnarray}\label{eq47}
\rho_{rh}=\sqrt{\frac{3\lambda}{3\lambda-2}}\,V_{end}\exp\Big[-3N_{rh}(1+\omega_{eff})\Big].
\end{eqnarray}
Now, by using equations (\ref{eq44}) and (\ref{eq47}) we can express
$\frac{a_{0}}{a_{rh}}$ in terms of $N_{rh}$ and $\omega_{eff}$ as
\begin{eqnarray}\label{eq48}
\ln\left(\frac{a_{0}}{a_{rh}}\right)=-\frac{1}{3}\ln\left(\frac{43}{11g_{rh}}\right)
-\frac{1}{4}\ln\left(\frac{\pi^{2}g_{rh}}{30\rho_{rh}}\right)-\ln
T_{0}
+\frac{1}{4}\ln\left(\sqrt{\frac{3\lambda}{3\lambda-2}}\,V_{end}\right)-\frac{3}{4}N_{rh}(1+\omega_{eff})\,.
\end{eqnarray}
From equations (\ref{eq15}) (in order to obtain $H_{hc}$),
(\ref{eq41}) and (\ref{eq48}), the e-folds number during reheating
is obtained as
\begin{eqnarray}\label{eq49}
N_{rh}=\frac{4}{1-3\omega_{eff}}\Bigg[-N-\ln\Big(\frac{k_{hc}}{a_{0}T_{0}}\Big)-\frac{1}{4}\ln\Big(\frac{40}{\pi^{2}g_{rh}}\Big)
-\frac{1}{3}\ln\Big(\frac{11g_{rh}}{43}\Big)+\frac{1}{2}\ln\Big(8\pi^{2}{\cal{A}}_{s}{\cal{W}}_{s}
c_{s}^{3}\Big)
\nonumber\\-\frac{1}{4}\ln\bigg(\sqrt{\frac{3\lambda}{3\lambda-2}}\,V_{end}\bigg)\Bigg].\hspace{1cm}
\end{eqnarray}
Also from equations (\ref{eq39}), (\ref{eq43}) and (\ref{eq46}) we
find the temperature during reheating as follows
\begin{eqnarray}\label{eq50}
T_{rh}=\bigg(\frac{30}{\pi^{2}g_{rh}}\bigg)^{\frac{1}{4}}\,
\bigg[\sqrt{\frac{3\lambda}{3\lambda-2}}\,V_{end}\bigg]^{\frac{1}{4}}
\times \exp\bigg[-\frac{3}{4}N_{rh}(1+\omega_{eff})\bigg]\,.
\end{eqnarray}
To study the reheating phase numerically, it is useful to rewrite
$N_{rh}$ and $T_{rh}$ in terms of the scalar spectral index. In this
respect, from equations (\ref{eq30}) and (\ref{eq35}) we have
\begin{eqnarray}\label{eq51}
\cos\bigg(\frac{\phi_{hc}}{f}\bigg)=-{\frac {{\it n_s}\,\beta+\sqrt
		{-6\,{\it n_s}\,\beta+6\,\beta+1}-\beta-1}{\beta \left( { \it n_s}-1
		\right) }}\nonumber\\=\frac{\sqrt{1+6\beta\delta_{ns}}-1}{\beta\delta_{ns}}-1 \,.\hspace{2.7cm}
\end{eqnarray}
By using equations (\ref{eq28}), (\ref{eq29}) and (\ref{eq51}) we
can write the number of e-folds during inflation ($N$) in terms of
$n_{s}$. In this way, we can obtain $N_{rh}$ and $T_{rh}$ as
functions of the scalar spectral index in order to seek for
numerical results. The results are shown in figures 4-7. Regarding
to the bound on the scalar spectral index from Planck2015 data, in
figure 4 we have plotted the ranges of $N_{rh}$ and $\omega_{eff}$
which makes the model observationally viable. In plotting the
figures we have used the sample values of $\beta$ as $\beta=12$ (the
left panel), $\beta=80$ (the middle panel) and $\beta=100$ (the
right panel). These values of $\beta$ are chosen based on the
constraints obtained in the previous section. Note also that we have
used the amplitude of the scalar power spectrum as
$A_{s}=2.196\times10^{-9}$. In the reheating analysis also, we can
obtain some constraints on $f$. This can be done by studying the
$r-n_{s}$ plane for several ranges of the effective equation of
state parameter during the reheating era and comparing the results
with Planck2015 data. The situation has been demonstrated in figure
5. Since it is likely that the value of the effective equation of
state parameter during the reheating phase to be in the range $0\leq
\omega_{eff}\leq \frac{1}{3}$, we can obtain some further
constraints on $f$ as well as the tensor-to-scalar ratio. Table 3
shows the constraints on $f$ and $r$. Our analysis shows that for
$\omega_{eff}=0$ and $\omega_{eff}=\frac{1}{3}$ there is no upper
limit on parameter $f$ (by considering the $95\%$ CL). Table 4 is
the same as table 3, but now by setting $\lambda\sim \Lambda^{-4}$
in order to state the obtained constraints just in terms of the reduced
Planck mass. Figure 6 shows the behavior of the number of e-folds
during reheating versus the scalar spectral index in confrontation
with Planck2015 data. To plot this figure we have adopted the sample
values $\beta=12,\,20,\,80$ and $100$. Our analysis shows that for
$20\leq\beta \leq 120$ the instantaneous reheating is favored by
Planck2015 data. In figure 7 we have plotted the temperature during
reheating versus the scalar spectral index for sample values of
$\beta$ as have been adopted in figure 6. Based on the bounds on the
scalar spectral index from Planck2015 data, we have obtained some
constraints on $N_{rh}$ and $T_{rh}$ summarized in tables 5 and 6.

\begin{tiny}
\begin{table*}
\caption{\label{tab3}Constraints on $f$ and $r$ based on the
Planck2015 data and by considering the values of the effective
equation of state parameter during the reheating era.}
\begin{tabular}{cccccccc}
\\ \hline \hline&$\omega_{eff}=0$&&$\omega_{eff}=\frac{1}{6}$
&&$\omega_{eff}=\frac{1}{3}$&&$\omega_{eff}=1$\\ \hline\\
$68\%$ CL& $---$
&&$\sqrt{\frac{16}{\lambda}}\frac{m_{pl}}{\Lambda^{2}}\leq f\leq
\sqrt{\frac{80}{\lambda}}\frac{m_{pl}}{\Lambda^{2}}$&&$\sqrt{\frac{13}{\lambda}}\frac{m_{pl}}{\Lambda^{2}}\leq
f\leq \sqrt{\frac{30}{\lambda}}\frac{m_{pl}}{\Lambda^{2}}$
&&$\sqrt{\frac{11}{\lambda}}\frac{m_{pl}}{\Lambda^{2}}< f\leq\sqrt{\frac{18}{\lambda}}\frac{m_{pl}}{\Lambda^{2}}$\\\\
$95\%$ CL& $\sqrt{\frac{18}{\lambda}}\frac{m_{pl}}{\Lambda^{2}}\leq
f$ &&$\sqrt{\frac{12}{\lambda}}\frac{m_{pl}}{\Lambda^{2}}<
f$&&$\sqrt{\frac{11}{\lambda}}\frac{m_{pl}}{\Lambda^{2}}< f\leq
\sqrt{\frac{100}{\lambda}}\frac{m_{pl}}{\Lambda^{2}}$
&&$\sqrt{\frac{10}{\lambda}}\frac{m_{pl}}{\Lambda^{2}}< f
<\sqrt{\frac{24}{\lambda}}\frac{m_{pl}}{\Lambda^{2}}$\\\\
\hline\\
$68\%$ CL&  $---$ &&$0.024\leq r \leq 0.052$&&$0.01\leq r\leq 0.027$
&&$0.0008\leq r \leq 0.001$\\\\
$95\%$ CL&  $0.039\leq r\leq 0.092$ &&$0.019\leq r \leq
0.075$&&$0.008\leq r\leq 0.058$
&&$0.0001\leq r \leq 0.002$\\\\
\hline
\end{tabular}
\end{table*}
\end{tiny}

\begin{small}
\begin{table*}
\caption{\label{tab4}Constraints on $f$ based on the Planck2015 data
by adopting $\lambda=\Lambda^{-4}$.}
\begin{tabular}{cccccccc}
\\ \hline \hline&$\omega_{eff}=0$&&$\omega_{eff}=\frac{1}{6}$
&&$\omega_{eff}=\frac{1}{3}$&&$\omega_{eff}=1$\\ \hline\\
$68\%$ CL& $---$ &&$\sqrt{16}\, m_{pl}\leq f\leq \sqrt{80}\,
m_{pl}$&&$\sqrt{13}\, m_{pl}\leq f\leq \sqrt{30}\, m_{pl}$
&&$\sqrt{11}\, m_{pl}< f\leq\sqrt{18}\, m_{pl}$\\\\
$95\%$ CL& $\sqrt{18}\, m_{pl}\leq f$ &&$\sqrt{12}\, m_{pl}\leq
f$&&$\sqrt{11}\, m_{pl}< f\leq \sqrt{100}\, m_{pl}$ &&$\sqrt{10}\,
m_{pl}< f <\sqrt{24}\, m_{pl}$\\\\
\hline
\end{tabular}
\end{table*}
\end{small}

\begin{figure}[ht]
\begin{center}
\includegraphics[scale=0.6]{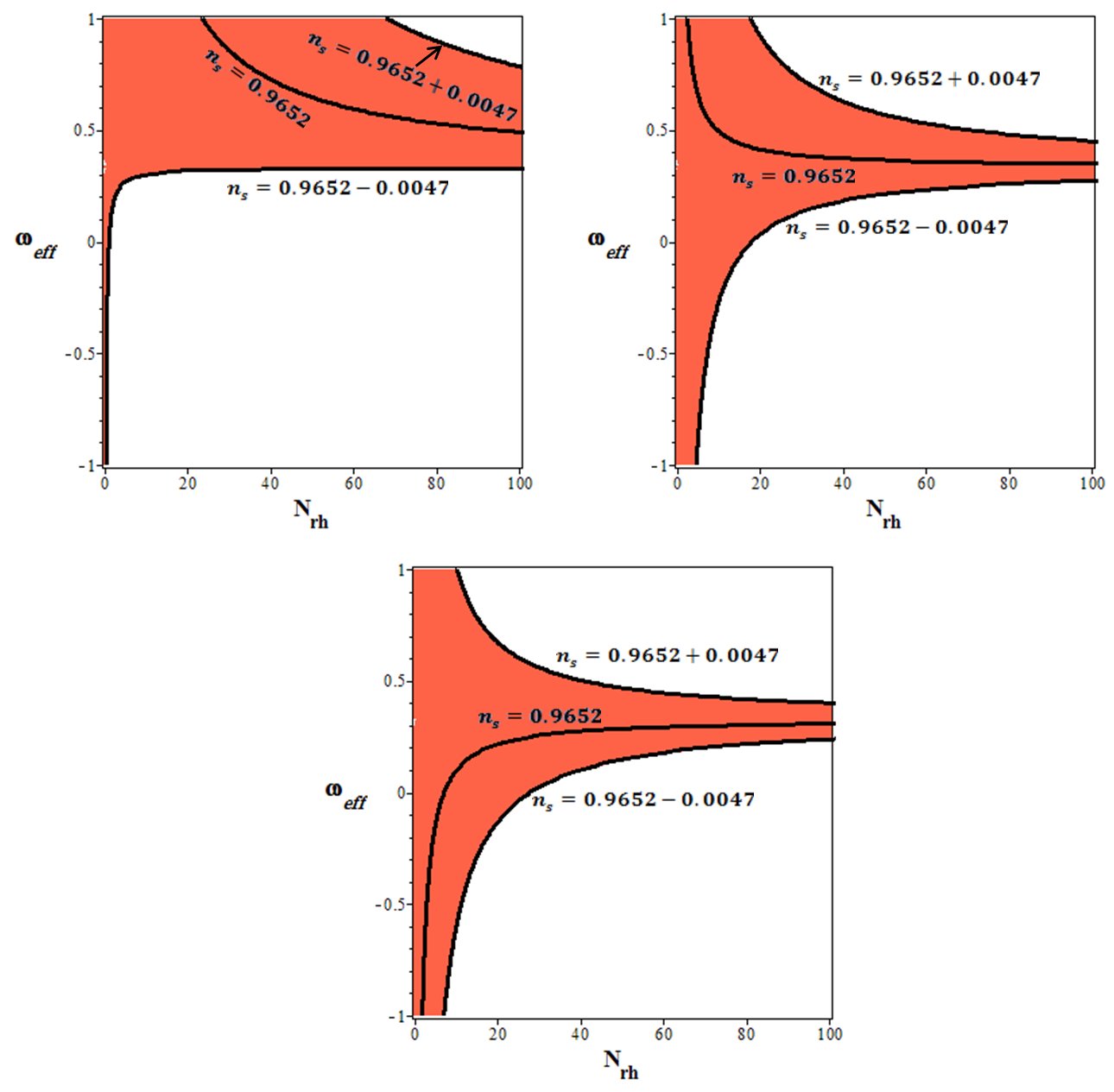}
\end{center}
\caption{\small {Observationally viable ranges of the
parameters $N_{rh}$ and $\omega_{eff}$ in our TNI model based on the
bounds on the scalar spectral index from Planck2015 data. The panels
are corresponding to $\beta=12$ (left panel) $\beta=80$ (middle
panel) and $\beta=100$ (right panel).}}
\label{fig3}
\end{figure}

\begin{figure}[ht]
\begin{center}
\includegraphics[scale=0.6]{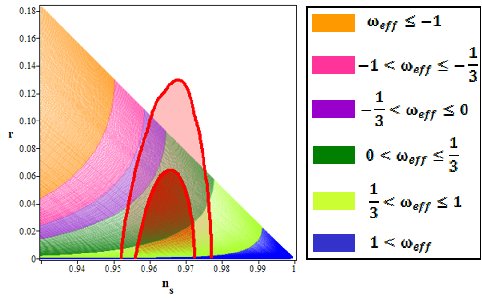}
\end{center}
\caption{\small {$r-n_{s}$ plane in TNI model for
several ranges of the effective equation of state parameter during
the reheating era in confrontation with Planck2015 TT, TE, EE+lowP
data.}}
\label{fig4}
\end{figure}

\begin{table}
\caption{\label{tab5} The ranges of the number of e-folds parameter
at reheating epoch in TNI model which are consistent with
observational data.}
\begin{tabular}{cccccccc}
\\ \hline \hline&$f=\sqrt{\frac{12}{\lambda}}\frac{m_{pl}}{\Lambda^{2}}$&&$f=\sqrt{\frac{20}{\lambda}}\frac{m_{pl}}{\Lambda^{2}}$&&
$f=\sqrt{\frac{80}{\lambda}}\frac{m_{pl}}{\Lambda^{2}}$&&$f=\sqrt{\frac{100}{\lambda}}\frac{m_{pl}}{\Lambda^{2}}$\\ \hline\\
$\omega=-1$&  $---$ &&$N_{rh}\leq0.19$&& $N_{rh}\leq4$
&&$N_{rh}\leq7$\\\\$\omega=-\frac{1}{3}$& $---$ &&$N_{rh}\leq0.32$&&
$N_{rh}\leq10$ &&$N_{rh}\leq14$\\\\$\omega=0$& $---$
&&$N_{rh}\leq0.61$&& $N_{rh}\leq19$ &&$N_{rh}\leq28$\\\\$\omega=1$&
$---$ &&$N_{rh}\leq69$&& $N_{rh}\leq18$
&&$N_{rh}\leq 10$\\\\
\hline
\end{tabular}
\end{table}

\begin{figure}[ht]
\begin{center}
\includegraphics[scale=0.5]{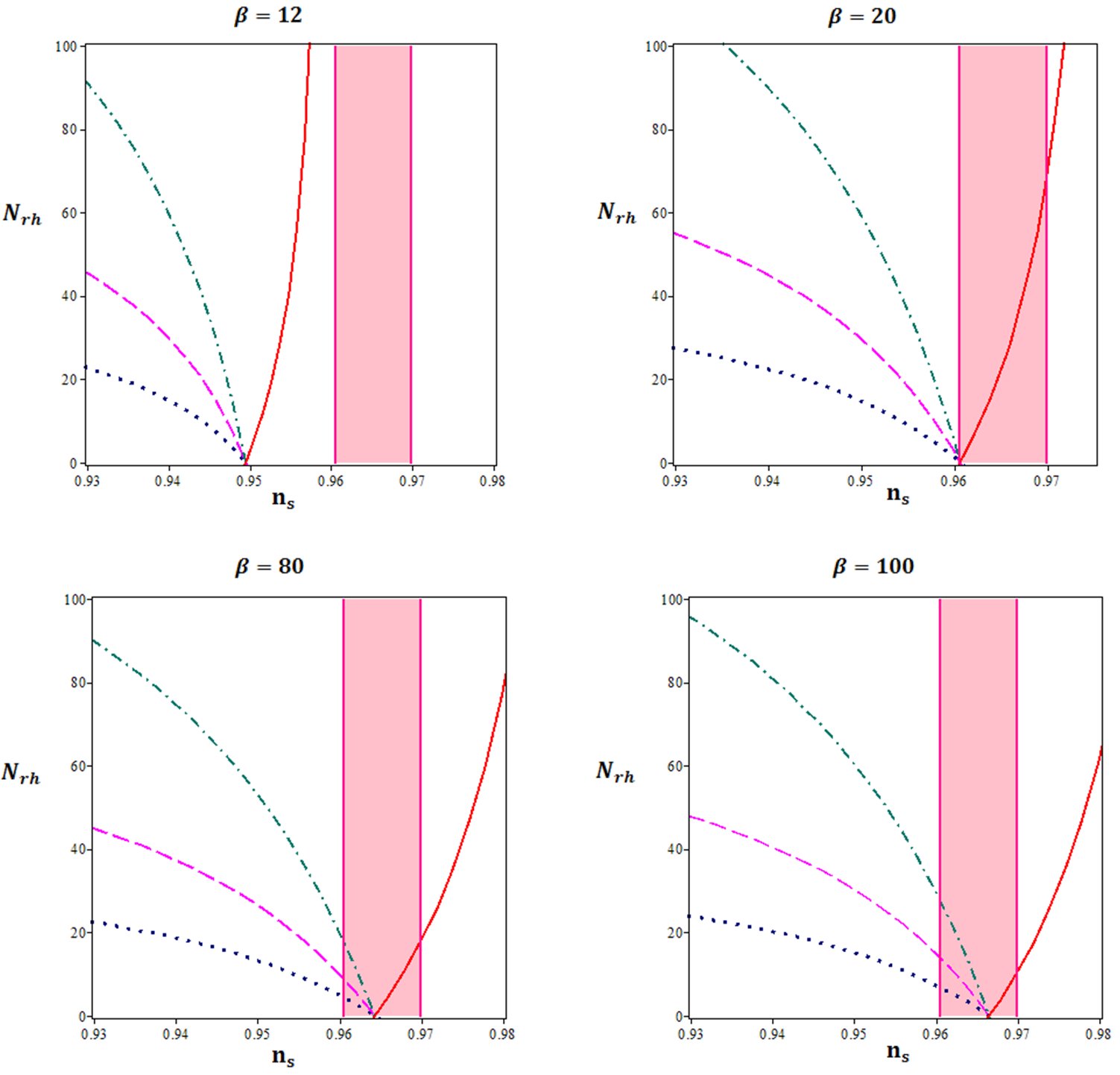}
\end{center}
\caption{\small {$N_{rh}$ versus $n_{s}$ for some sample
values of $\beta$ in confrontation with Planck2015 data. The values
of the effective equation of state parameter are chosen as
$\omega_{eff}=-1$ (dotted line), $\omega_{eff}=-\frac{1}{3}$ (dashed
line), $\omega_{eff}=0$ (dashed-dotted line) and $\omega_{eff}=1$
(solid line). The pink region is the bound on $n_{s}$ from
Planck2015.}}
\label{fig5}
\end{figure}

\begin{figure}[ht]
\begin{center}
\includegraphics[scale=0.5]{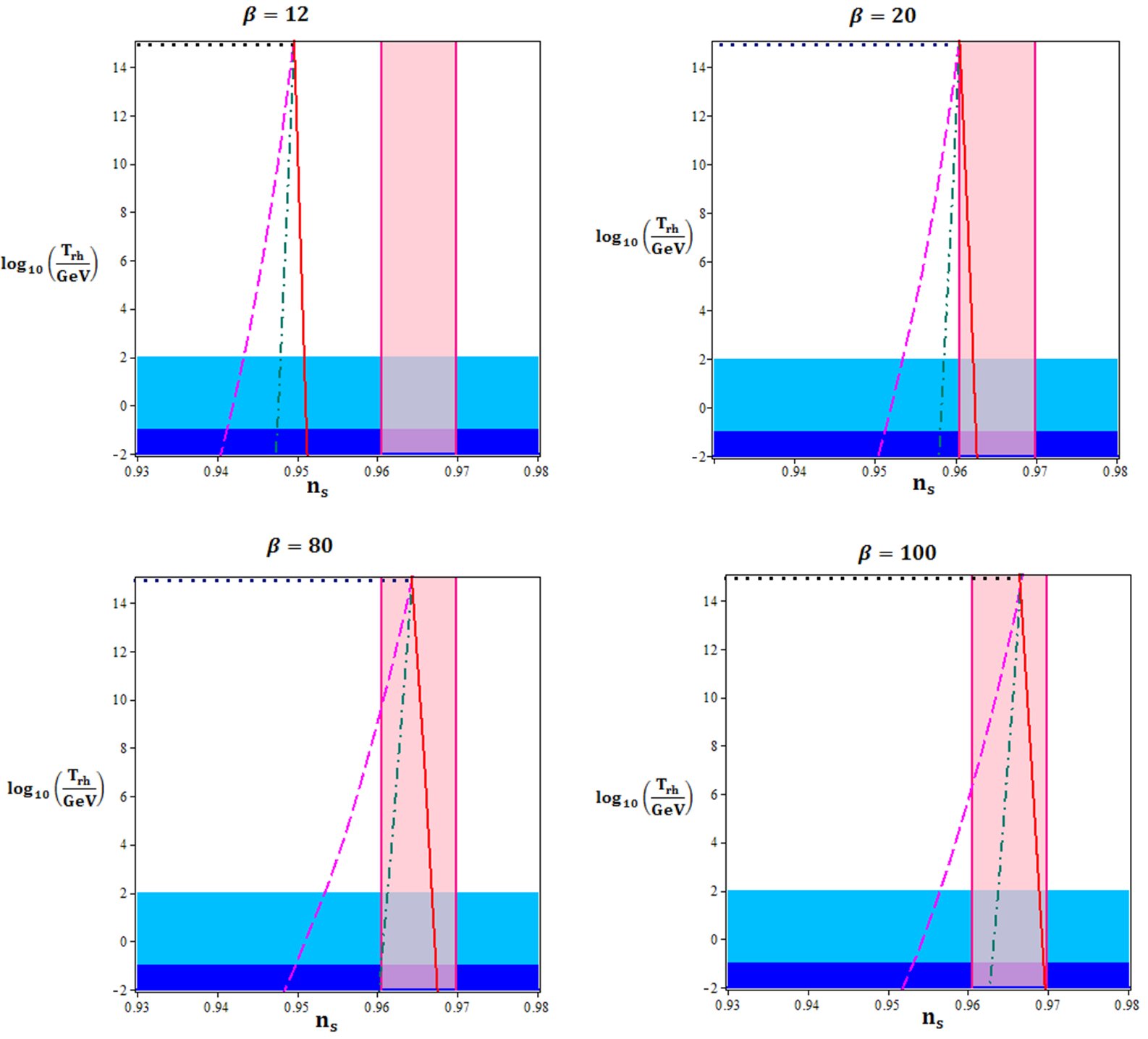}
\end{center}
\caption{\small {$\log_{10}\left(\frac{T_{rh}}{GeV}\right)$ versus $n_{s}$ for some sample values of $\beta$ and $\omega_{eff}$ in confrontation with
Planck2015 data. The values of the effective equation of state
parameter are chosen as $\omega_{eff}=-1$ (dotted line),
$\omega_{eff}=-\frac{1}{3}$ (dashed line), $\omega_{eff}=0$
(dashed-dotted line) and $\omega_{eff}=1$ (solid line). The light
blue region demonstrates the temperatures below the Electroweak
scale, $T<100$ GeV, and the dark blue region shows the temperatures
below the Big Bang Nucleosynthesis scale, $T<10$ MeV.}}
\label{fig6}
\end{figure}

\begin{table*}
\caption{\label{tab6} The ranges of the temperature at reheating
epoch in our TNI model which are consistent with observational
data.}
\begin{tabular}{cccccccc}
\\ \hline \hline&$f=\sqrt{\frac{12}{\lambda}}\frac{m_{pl}}{\Lambda^{2}}$&&$f=\sqrt{\frac{20}{\lambda}}\frac{m_{pl}}{\Lambda^{2}}$&&
$f=\sqrt{\frac{80}{\lambda}}\frac{m_{pl}}{\Lambda^{2}}$&&$f=\sqrt{\frac{100}{\lambda}}\frac{m_{pl}}{\Lambda^{2}}$\\ \hline\\
$\omega=-\frac{1}{3}$& $---$
&&$\log_{10}\left(\frac{T_{rh}}{GeV}\right)\geq14.9$&&
$\log_{10}\left(\frac{T_{rh}}{GeV}\right)\geq9.5$
&&$\log_{10}\left(\frac{T_{rh}}{GeV}\right)\leq 6.82$\\\\$\omega=0$&
$---$ &&$\log_{10}\left(\frac{T_{rh}}{GeV}\right)\geq14.5$&&
$\log_{10}\left(\frac{T_{rh}}{GeV}\right)\geq-1.3$
&&$---$\\\\$\omega=1$& $---$ &&$---$&& $---$
&&$---$\\\\
\hline
\end{tabular}
\end{table*}

\section{Summary}

By considering a tachyon field to be responsible for Natural
Inflation, we have constructed a Tachyon Natural Inflation. Tachyon
Natural Inflation with large values of parameter $f$ (the width of the
potential) meets the tachyon model with $\phi^{2}$ potential which
is consistent with Planck2015 observational data. We have studied
the cosmic inflation and linear perturbations in this setup. In this
way, we have obtained the slow-roll parameters ($\epsilon$, $\eta$
and $\zeta$) and the main perturbation parameters such as the scalar
spectral index, its running and the tensor-to-scalar ratio. Because
of the form of the Lagrangian in this setup, these parameters have
been obtained in terms of the height of potential ($2\Lambda^{4}$),
its width ($f$), reduced Planck mass ($m_{pl}$) and the warp factor
($\lambda$). As we know, in canonical Natural Inflation (NI) these
are obtained in terms of only the width of the potential and the reduced Planck
mass. We have performed a numerical analysis on the model parameter
space and compared the results with $68\%$ CL and $95\%$ CL regions
of Planck2015 data. In this regard and by studying $r-n_{s}$ and
$\alpha_{s}-n_{s}$ planes we have obtained some constraints on
parameter $f$, which is in terms of $\lambda$, $\Lambda$ and
$m_{pl}$. We have performed our analysis by adopting $N=50$, $N=55$
and $N=60$. By analyzing the perturbation parameters we have found
lower limits on $f$, however there is no upper limits on this
parameter in confrontation with $95\%$ CL of Planck2015 data. Note
that, if we consider $\lambda\sim\Lambda^{-2}$, the obtained
constraints satisfy the condition (\ref{eq1}). As figure 1 shows,
our TNI model has very good agreement with Planck2015 data, much
better than the canonical natural inflation. We have also shown that
a prediction of this model is $r\leq\frac{8}{3}\,\delta_{ns}$, where
$\delta_{ns}$ is the scalar spectral tilt. Considering that from the
Planck2015 data we have $\delta_{ns}=0.032$, we get the constraint
$r\leq 0.085$. The reheating era after inflation has been studied
also in this NTI model. We have obtained the number of e-folds and
temperature during reheating era in terms of $n_{s}$, $\beta$ and
$\omega_{eff}$. Taking into account that it is likely for the
value of the effective equation of state parameter to be between $0$
and $\frac{1}{3}$, we have obtained some explicit constraints on the
width of potential and the tensor-to-scalar ratio in this setup.
These constraints are obtained based on the observationally viable
values of the scalar spectral index from $68\%$ CL and $95\%$ CL
regions of Planck2015 data. We have also studied the number of e-folds
and temperature during the reheating era numerically. Regarding the
bounds on the scalar spectral index from Planck2015 data set, we
have obtained new constraints on $N_{rh}$ and $T_{rh}$.

In summary, the Tachyon Natural Inflation model presented in this paper is a
cosmologically viable model (much better than the standard canonical
natural inflation) which its perturbation parameters values lie well
within the $95\%$ (and $68\%$) CL region of the Planck2015 dataset,
even in large $f$ limit. Also, the effective equation of state
parameter in this model for $0\leq \omega_{eff} \leq \frac{1}{3}$
gives the observationally viable values of the scalar spectral index and tensor-to-scalar ratio.

{\bf Acknowledgement}\\
The work of K. Nozari has been supported financially by Research
Institute for Astronomy and Astrophysics of Maragha (RIAAM) under
research project
number 1/5237-10.\\

\end{document}